# Coupling the Minkowski's theory with the Maxwell's equations for a mechano-driven media system for engineering electromagnetism

Zhong Lin Wang*

Beijing Institute of Nanoenergy and Nanosystems, Chinese Academy of Sciences, Beijing 101400, P. R. China
*e-mail: zhong.wang@mse.gatech.edu

**Abstract**

The Minkowski's theory is regarded as the classical approach for describing the electromagnetism of uniformly moving objects by elegantly utilizing the format-invariance of the Maxwell's equations in inertia reference frames under Lorentz transformation. However, in practice, we usually have the cases of a medium that has boundary and moves with acceleration, such as a rotating dielectric disc or a magnetic object. To fully describe the electromagnetic behavior of such a system, we have developed the *Maxwell's equations for a mechano-driven media system* (MEs-f-MDMS). This paper is to expand the Minkowski's approach for deriving the constitutive equations and the boundary conditions for MEs-f-MDMS under low-speed approximation ($v \ll c$), based on which the equations can be properly solved for describing engineering electromagnetism, with considering the coupling among electric field - magnetic field - mechanical force field. In practice, mechanical action is not merely passive motion through space but an active source of electromagnetic generation and modulation. Our theory makes it possible to treat the electrodynamics by including medium deformation, strain gradients, rotational dynamics, and interfacial contact processes in the full calculation.

## 1. Introduction

The theoretical foundation of the classical electrodynamics was first established by Maxwell in 1865 [1], and later on extensively developed and reformulated by Heaviside around 1900 [2]. The equations unified electric and magnetic phenomena into a self-consistent field theory, forming the foundation for wireless communication, signal transmission, radar, photonics and many more. In material media, Maxwell's equations must be complemented by constitutive relations that describe how matter responds to electromagnetic fields. While the conventional treatment assumes stationary or quasi-static media, many physical systems involve motion, deformation, and mechanical triggering that fundamentally alter electromagnetic behavior, especially if one is interested in near-field electromagnetic behavior.

A major step toward a relativistically consistent description of electrodynamics in moving matter was first developed by Minkowski in 1908 [3]. Minkowski reformulated macroscopic electrodynamics within four-dimensional space-time, embedding electromagnetic fields and material responses into a covariant framework. In this theory, the motion of the medium directly influences the electromagnetic excitation fields because the velocity of the medium enters intrinsically into the constitutive equations. Minkowski's formulation provides the conceptual starting point for understanding electrodynamics in moving continua [4,5]. However, it primarily addresses uniform translational motion and does not fully incorporate internally driven mechanical processes such as strain, stress, rotation, dynamic contact, or structural deformation. In modern mechano-driven media system, mechanical action is not merely passive motion through space but an active source of electromagnetic generation and modulation. Mechanical degrees of freedom — including deformation fields, strain gradients, rotational dynamics, and interfacial contact processes — can directly induce polarization, magnetization, charge redistribution, and current generation.

The macroscopic and continuum treatment of electromagnetic fields in matter was systematically presented by Landau and Lifshitz [6], which remains the standard reference for moving dielectrics, stress tensors, and boundary conditions. Modern clarifications of electromagnetic momentum and force in moving media were further advanced by Obukhov in 2008 [7]. Extensions to accelerated and rotating systems, directly relevant to mechano-driven media, were analyzed in the covariant treatment of inhomogeneous rotating media by Goto, Tucker, and Walton in 2011 [8]. In addition, the coupling between electromagnetic fields and mechanical continua has been systematically examined in the relativistic continuum framework by de Groot and Suttorp [9], establishing the theoretical basis for incorporating deformation, strain, and stress into electromagnetic field equations. Ridgely used covariant field equations to derive general field equations in a rotating coordinate system [10]. Constitutive equations are then derived in the rotating and laboratory reference frames. Together, these works form the essential theoretical background upon which the expanded Minkowski approach for mechano-driven media systems is constructed. These approaches mainly use Minkowski's approach without make any expansion to the Maxwell's equations event when the dielectric medium has an accelerated motion.

Recently, the theory of Maxwell's equations for a mechano-driven media system (MEs-f-MDMS) was proposed by Wang to systematically incorporate mechanical state variables into the electromagnetic framework [11-13]. In this generalized description, the material response depends not only on electromagnetic field variables and bulk velocity, but also on non-equilibrium mechanical driving (accelerated motion). Mechanical excitation is therefore elevated to the same theoretical status as electromagnetic excitation, leading to a unified mechano-electromagnetic field theory. As a result of three fields coupling, the MEs-f-MDMS is not format-invariant under Lorentz transformation.

In this paper, by utilizing the Minkowski approach, we have derived the constitutive equations and the associated boundary conditions for the MEs-f-MDMS, providing a rigorous theoretical system for describing electromagnetic phenomena in dynamically driven media, so that the theory is complete and ready to be utilized for numerical calculations. Such a formulation not only advances the fundamental understanding of electrodynamics in non-equilibrium matter but also establishes the theoretical foundation for technologies based on mechanically driven electromagnetic processes, including deformable dielectrics, triboelectric systems, and other dynamic energy conversion platforms.

## 2. Lorentz transformation for treating the electromagnetism of point charges in free-space

In classical electrodynamics, since the Maxwell's equations were first derived for stationary medium system, the electromagnetism for a moving observer in free-space is thus treated using Lorentz transformation by assuming that the format-invariance of the Maxwell's equations, which means that the format of the Maxwell's equations preserves in two inertia reference frames that have a relative uniform movement. This is the basic idea for dealing with the electromagnetic behavior in moving reference frame with considering only the electric-magnetic coupling [14].

The electromagnetic behavior of a moving observer is rather easy to be done in free-space, where there is no finite size objects but distributed point charges, so no boundary conditions need to be considered except at infinity. In Lorentz transformation, the invariance of speed of light in free-space is assumed. Therefore, the electromagnetic fields observed in a moving frame ($S'$) can be derived from that in the Lab reference frame ($S$) by preserving the format-invariance of the Maxwell equations with the use of Lorentz transformation. Therefore, the electromagnetic phenomenon of a point charge in free-space observed by two independent observers located in two inertial reference frames that have a relative movement at a constant speed. Now we assume that an observer named Alice is in a moving inertial frame $S'$ that moves at a velocity $\boldsymbol{v_0}$ relative to the non-moving frame $S$ (Lab frame), as shown in Fig. 1. If there is a point charge $+q$ that is at rest in the Lab frame $S$, what Bob in the Lab frame observes is only a Coulomb field. As observed by Alice in the moving frame $S'$, the point charge is moving at a relative velocity of $-\boldsymbol{v}_0$, so that she will detect not only an electric field but also a magnetic field as caused by the moving charge $+q$. The two relatively moving observers see different electromagnetic behaviors although the event is the same. To correlate the two sets of fields observed by Bob and Alice, coordination transformation is usually adopted. For simplicity, we assume the moving direction is along the x-axis, the Lorentz transformation is given as:

$$x' = \gamma(x - v_0 t) \quad (1a)$$

$$y' = y \quad (1b)$$

$z' = z$ (1c)

$t' = \gamma(t - xv_0/c^2)$ (1d)

where

$\gamma = 1/(1 - v_0^2/c^2)^{1/2},$ (2a)

and $c = 1/\sqrt{\varepsilon_0 \mu_0}$ (2b)

is the speed of light in free-space, where $\varepsilon_0$ and $\mu_0$ are the electric and magnetic permittivity, respectively, in free-space.

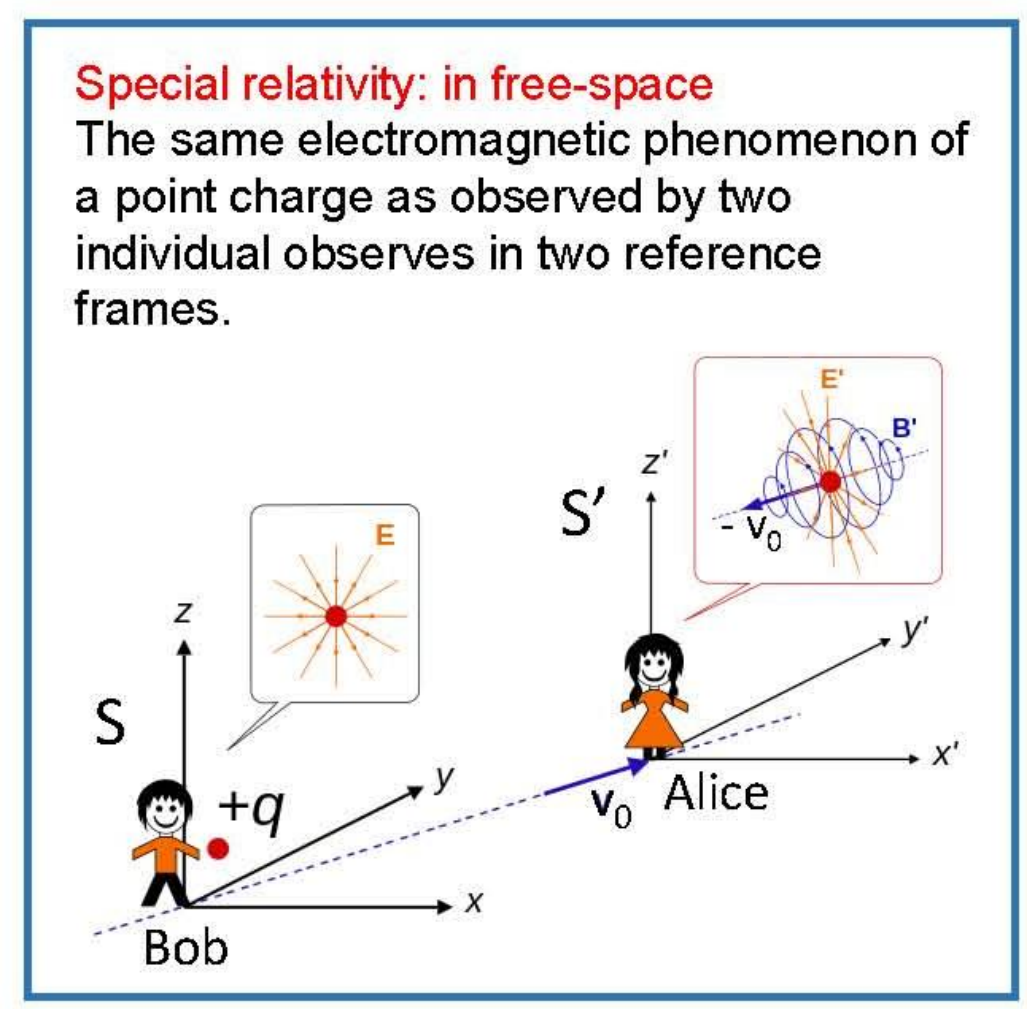

Figure 1. Classical approach for treating the electromagnetic behavior of point charges in free-space as observed by two observers who have relative uniform motion using Lorentz transformation. This is the basis of special relativity.

## 3. Electrodynamics of a uniformly moving object – the Minkowski theory

### 3.1 General approach

We now consider a case that an observer is in the Lab frame, a medium/object with finite size is moving at a uniform velocity (Fig. 2). In the moving frame that is affixed to the medium in the $S'$ frame, in which the medium is at stationary, so the local fields satisfy the classical Maxwell's equations:

$\nabla' \cdot \boldsymbol{D}' = \rho'$ (3a)

$\nabla' \cdot \boldsymbol{B}' = 0$ (3b)

$$\nabla' \times \boldsymbol{E}' = -\frac{\partial}{\partial t'}\boldsymbol{B}' \tag{3c}$$

$$\nabla' \times \boldsymbol{H}' = \boldsymbol{J}' + \frac{\partial}{\partial t'}\boldsymbol{D}' \tag{3d}$$

The constitutive equations for a stationary medium case are:

$$\boldsymbol{B}' = \mu \boldsymbol{H}' \tag{4a}$$

$$\boldsymbol{D}' = \varepsilon \boldsymbol{E}' \tag{4b}$$

where $\varepsilon$ and μ are the electric and magnetic permittivity, respectively, of the dielectric medium, which are assumed to be constant and independent of the motion of the medium.

In the Lab frame, the electromagnetic behavior is also assumed to be governed by the Maxwell equations if the moving velocity of the medium is a constant:

$$\nabla \cdot \boldsymbol{D} = \rho \tag{5a}$$

$$\nabla \cdot \boldsymbol{B} = 0 \tag{5b}$$

$$\nabla \times \boldsymbol{E} = -\frac{\partial}{\partial t}\boldsymbol{B} \tag{5c}$$

$$\nabla \times \boldsymbol{H} = \boldsymbol{J} + \frac{\partial}{\partial t}\boldsymbol{D} \tag{5d}$$

This means that the format-invariance of the Maxwell equations in inertia reference frames was also assumed in Minkowski theory.

In general, the electromagnetic behavior observed in the Lab frame $(\boldsymbol{r}, t)$ $(\boldsymbol{E}, \boldsymbol{B}, \boldsymbol{D}, \boldsymbol{H})$ and those fields in the moving frame affixed to the medium $(\boldsymbol{r}', t')$ $(\boldsymbol{E}', \boldsymbol{B}', \boldsymbol{D}', \boldsymbol{H}')$ are correlated by applying the Lorentz transformation to the Maxwell equations in the moving frame (Eqs. (3a-d)) and construct the final expression into the format of the Maxwell equations in the Lab frame (Eqs. (5a-d). For simplicity, we assume the moving direction is along the x-axis, the Lorentz transformation is

$$\frac{\partial}{\partial t'} = \gamma\left(\frac{\partial}{\partial t} + v_0 \frac{\partial}{\partial x}\right) \tag{6a}$$

$$\frac{\partial}{\partial x'} = \gamma\left(\frac{\partial}{\partial x} + v_0/c^2 \frac{\partial}{\partial t}\right) \tag{6b}$$

$$\frac{\partial}{\partial y'} = \frac{\partial}{\partial y} \tag{6c}$$

$$\frac{\partial}{\partial z'} = \frac{\partial}{\partial z} \tag{6d}$$

After some mathematical work on the four equations, the fields in parallel and in perpendicular to the moving direction of the reference frame are correlated by:

$$\boldsymbol{E}'_{//} = \boldsymbol{E}_{//},\ \boldsymbol{B}'_{//} = \boldsymbol{B}_{//},\ \boldsymbol{D}'_{//} = \boldsymbol{D}_{//},\ \boldsymbol{H}'_{//} = \boldsymbol{H}_{//}, \tag{7a}$$

$$\boldsymbol{E}'_{\perp} = \gamma(\boldsymbol{E} + \boldsymbol{v}_0 \times \boldsymbol{B})_{\perp} \tag{7b}$$

$$\boldsymbol{B}'_{\perp} = \gamma(\boldsymbol{B} - \ \boldsymbol{v}_0 \times \boldsymbol{E}/c^2)_{\perp} \tag{7c}$$

$$\boldsymbol{D}'_{\perp} = \gamma(\boldsymbol{D} + \boldsymbol{v}_0 \times \boldsymbol{H}/c^2)_{\perp} \tag{7d}$$

$$\boldsymbol{H}'_{\perp} = \gamma(\boldsymbol{H} \ - \boldsymbol{v}_0 \times \boldsymbol{D})_{\perp} \tag{7e}$$

$$\boldsymbol{J}'_{//} = \gamma(\boldsymbol{J}_{//} \ - \rho \boldsymbol{v}_0) \tag{7f}$$

$$\boldsymbol{J}'_{\perp} = \boldsymbol{J}_{\perp} \tag{7g}$$

$$\rho' = \gamma(\rho - \boldsymbol{v}_0 \cdot \boldsymbol{J}/c^2) \tag{7h}$$

It needs to point out that the most amazing and beauty point here is that the Maxwell's equations can be truly made into format-invariant under the Lorentz transformation in free-space, just as originally assumed by Minkowski. This is the magic of Maxwell equations and the Lorentz transformation if the object is moving at a constant velocity $\boldsymbol{v}_0$!

The above theory was first proposed by Minkowski for a moving object that moves with uniform velocity by assuming that the format-invariance of the Maxwell's equations. Because the fields satisfy the superposition principle, the fields generated by several moving objects can be solved individually and then add up together, provided the moving velocity of each is uniform.

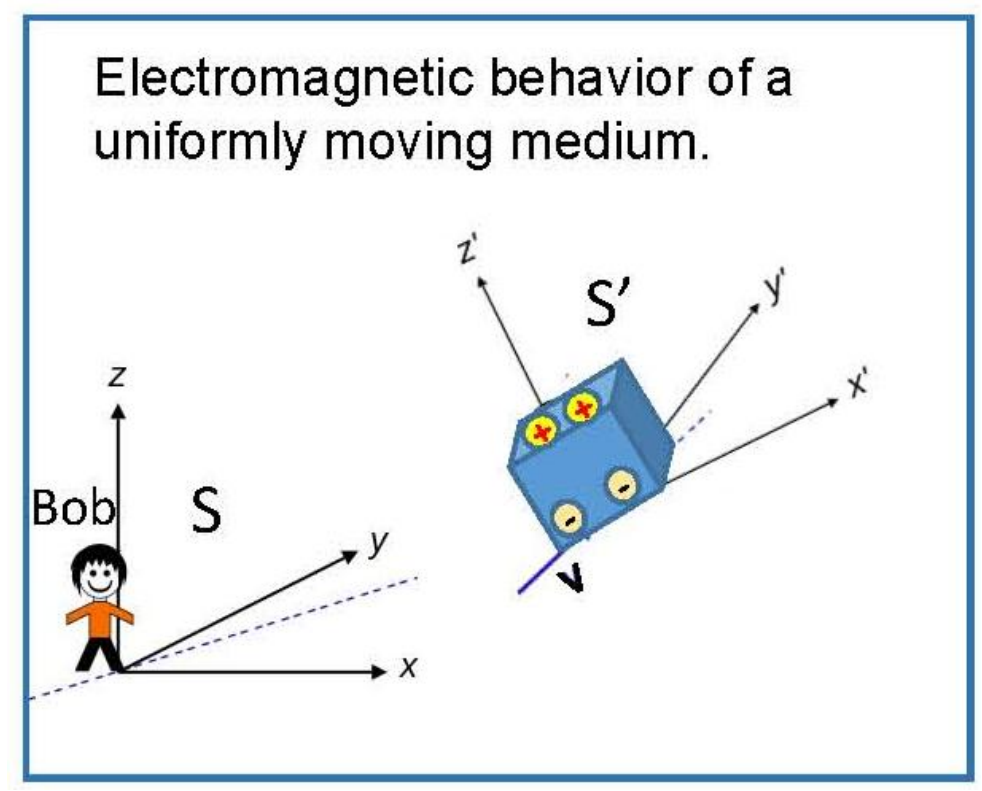

Figure 2. Schematic diagram showing the coordination system for treating the electrodynamics of a uniformly moving dielectric medium in two inertia reference frames.

However, for an object that moves with acceleration, such as a rotating disc. The format-invariance of the Maxwell's equations may not hold, as first pointed out by Feynman in his famous "anti-flux" examples [15]. Such discussions will be presented in Section 4.

### 3.2 The constitutive equations

In electromagnetic theory, we have four quantities: electric field $\boldsymbol{E}$, electric displacement vector $\boldsymbol{D}$, magnetic field $\boldsymbol{H}$, and magnetic induction $\boldsymbol{B}$, the equations that govern their relationships are called the constitutive equations. We now derive the constitutive equations for a moving medium case. Eqs. (7a-d) can be more generally written as [16]:

$$\boldsymbol{E}' = \gamma\left[\boldsymbol{E} - \frac{(\gamma-1)}{\gamma}\frac{\boldsymbol{v}_0(\boldsymbol{v}_0\cdot\boldsymbol{E})}{{v_0}^2} + \boldsymbol{v}_0\times\boldsymbol{B}\right] \tag{8b}$$

$$\boldsymbol{B}' = \gamma\left[\boldsymbol{B} - \frac{(\gamma-1)}{\gamma}\frac{\boldsymbol{v}_0(\boldsymbol{v}_0\cdot\boldsymbol{B})}{{v_0}^2} - \frac{1}{c^2}\boldsymbol{v}_0\times\boldsymbol{E}\right] \tag{8b}$$

$$\boldsymbol{D}' = \gamma\left[\boldsymbol{D} - \frac{(\gamma-1)}{\gamma}\frac{\boldsymbol{v}_0(\boldsymbol{v}_0\cdot\boldsymbol{D})}{{v_0}^2} + \frac{1}{c^2}\boldsymbol{v}_0\times\boldsymbol{H}\right] \tag{8b}$$

$$\boldsymbol{H}' = \gamma\left[\boldsymbol{H} - \frac{(\gamma-1)}{\gamma}\frac{\boldsymbol{v}_0(\boldsymbol{v}_0\cdot\boldsymbol{H})}{{v_0}^2} - \boldsymbol{v}_0\times\boldsymbol{D}\right] \tag{8b}$$

Substituting Eqs. (8a-d) into the constitutive equations (4a-b) in the moving frame:

$$\boldsymbol{D} - \frac{(\gamma-1)}{\gamma}\frac{\boldsymbol{v}_0(\boldsymbol{v}_0\cdot\boldsymbol{D})}{{v_0}^2} + \frac{1}{c^2}\boldsymbol{v}_0\times\boldsymbol{H} = \varepsilon\left[\boldsymbol{E} - \frac{(\gamma-1)}{\gamma}\frac{\boldsymbol{v}_0(\boldsymbol{v}_0\cdot\boldsymbol{E})}{{v_0}^2} + \boldsymbol{v}_0\times\boldsymbol{B}\right] \tag{9a}$$

$$\boldsymbol{B} - \frac{(\gamma-1)}{\gamma}\frac{\boldsymbol{v}_0(\boldsymbol{v}_0\cdot\boldsymbol{B})}{{v_0}^2} - \frac{1}{c^2}\boldsymbol{v}_0\times\boldsymbol{E} = \mu\left[\boldsymbol{H} - \frac{(\gamma-1)}{\gamma}\frac{\boldsymbol{v}_0(\boldsymbol{v}_0\cdot\boldsymbol{H})}{{v_0}^2} - \boldsymbol{v}_0\times\boldsymbol{D}\right] \tag{9b}$$

Making a scalar product of $\boldsymbol{v}_0$ on Eqs. (9a-b), we have

$$\boldsymbol{v}_0\cdot\boldsymbol{D} = \varepsilon\,\boldsymbol{v}_0\cdot\boldsymbol{E} \tag{10a}$$

$$\boldsymbol{v}_0\cdot\boldsymbol{B} = \mu\,\boldsymbol{v}_0\cdot\boldsymbol{H} \tag{10b}$$

Using Eqs. (10a-b), Eqs. (9a-b) are simplified as

$$\boldsymbol{D} + \frac{1}{c^2}\boldsymbol{v}_0\times\boldsymbol{H} = \varepsilon(\boldsymbol{E} + \boldsymbol{v}_0\times\boldsymbol{B}) \tag{11a}$$

$$\boldsymbol{B} - \frac{1}{c^2}\boldsymbol{v}_0\times\boldsymbol{E} = \mu(\boldsymbol{H} - \boldsymbol{v}_0\times\boldsymbol{D}) \tag{11b}$$

By making a cross of $\boldsymbol{v}_0\times$ on Eqs. (11a-b), and using a vector calculation formula

$\boldsymbol{v}_0 \times (\boldsymbol{v}_0 \times \boldsymbol{a}) = \boldsymbol{v}_0(\boldsymbol{v}_0 \cdot \boldsymbol{a}) - v_0{}^2 \boldsymbol{a}$,

for a general vector $\boldsymbol{a}$, the constitutive equations can be received [17]:

$$\boldsymbol{D} = \frac{1}{1-\varepsilon\mu v_0{}^2}\{(1 - \frac{v_0{}^2}{c^2})\varepsilon\boldsymbol{E} + (\varepsilon\mu - \frac{1}{c^2})[\boldsymbol{v}_0 \times \boldsymbol{H} - \varepsilon\boldsymbol{v}_0(\boldsymbol{v}_0 \cdot \boldsymbol{E})]\} \quad (12a)$$

$$\boldsymbol{B} = \frac{1}{1-\varepsilon\mu v_0{}^2}\{(1 - \frac{v_0{}^2}{c^2})\mu\boldsymbol{H} - (\varepsilon\mu - \frac{1}{c^2})[\boldsymbol{v}_0 \times \boldsymbol{E} - \mu\boldsymbol{v}_0(\boldsymbol{v}_0 \cdot \boldsymbol{H})]\} \quad (12b)$$

One is interested to note that there is a factor of $(\varepsilon\mu - \frac{1}{c^2}) = (\frac{1}{c_m^2} - \frac{1}{c^2})$ appearing in the equation. This means that both speeds of light in free-space and in medium appear in the same equation! This is a consequence of using the Lorentz transformation developed for free-space to treat the electromagnetic behavior in a moving dielectric medium case. We believe that this may be a remaining question for the Minkowski theory to be solved.

### 3.3 Low-speed approximated results

To make discussions simple, we now consider the results for low-speed approximation, $v_0 \ll c$, so that $\gamma \approx 1$, from Eqs. (8a-d), the fields in the two reference frames are related approximately by [17]:

$$\boldsymbol{E}' \approx \boldsymbol{E} + \boldsymbol{v}_0 \times \boldsymbol{B} \quad (13a)$$

$$\boldsymbol{B}' \approx \boldsymbol{B} - \boldsymbol{v}_0 \times \boldsymbol{E}/c^2 \quad (13b)$$

$$\boldsymbol{D}' \approx \boldsymbol{D} + \boldsymbol{v}_0 \times \boldsymbol{H}/c^2 \quad (13c)$$

$$\boldsymbol{H}' \approx \boldsymbol{H} - \boldsymbol{v}_0 \times \boldsymbol{D} \quad (13d)$$

$$\boldsymbol{J}' \approx \boldsymbol{J} - \rho\boldsymbol{v}_0 \quad (13e)$$

$$\rho' \approx \rho - \boldsymbol{v}_0 \cdot \boldsymbol{J}/c^2 \quad (13f)$$

The corresponding constitutive equations are:

$$\boldsymbol{D} \approx \varepsilon\boldsymbol{E} + \alpha\varepsilon\mu[\boldsymbol{v}_0 \times \boldsymbol{H} - \varepsilon\boldsymbol{v}_0(\boldsymbol{v}_0 \cdot \boldsymbol{E})] \quad (14a)$$

$$\boldsymbol{B} = \mu\boldsymbol{H} - \alpha\varepsilon\mu[\boldsymbol{v}_0 \times \boldsymbol{E} - \mu\boldsymbol{v}_0(\boldsymbol{v}_0 \cdot \boldsymbol{H})] \quad (14b)$$

where we define a factor:

$$\alpha = 1 - \frac{\varepsilon_0\mu_0}{\varepsilon\mu} = 1 - c^2/c_m{}^2 \quad (14c)$$

## 4. Deriving the Maxwell's equations for a mechano-driven medium system (MEs-f-MDMS)

### 4.1 Galilean space and time

All of the original theory for physics was first derived based on the Galilean transformation by assuming that time and space are independent:

$$x' = x - v_0 t \tag{15a}$$

$$y' = y \tag{15b}$$

$$z' = z \tag{15c}$$

$$t' = t \tag{15d}$$

Later after Einstein, Lorentz transformation is used to derive the electromagnetic phenomenon in general. It appears that the Galilean transformation and Lorentz transformation are inconsistent because they represent the classical and new space-time relationships, respectively. However, it has been proved that the Lorentz transformation is equivalent to Galilean transformation under two approximations [18,19]:

1. The relative speed between two inertial frames of reference is much smaller than the speed of light in vacuum: $v_0 \ll c$, so $\gamma \approx 1$; and
2. Electromagnetic phenomenon takes place in an arena, the spatial extension of which is much smaller than the distance traveled by light during the duration of the phenomenon: $x \ll ct$. From Eq. (1d), under the approximation of $v_0 \ll c$, the Lorentz transformation for time becomes $t' \approx \left(t - \frac{xv_0}{c^2}\right) = \frac{1}{c}\left(ct - x\frac{v_0}{c}\right)$. If $x \ll ct$, the space caused correction to time is so small and it can completely ignored, resulting $t' \approx t$, which is the result of Galilean transformation.

These two conditions are usually met in engineering electrodynamics, so that Galilean transformation can be safely used.

### 4.2 Expansion of the Minkowski theory to low-speed moving object with acceleration

We now consider a case that an observer is in the Lab frame, a medium/object with finite size is at accelerated motion, which occurs a lot in engineering electromagnetism, such as power generator, wind blades and extremely low-frequency electromagnetic wave generation. As presented in Fig. 3. We assume that the object can have an accelerated motion, but its motion within a small time interval $\Delta t$ can be viewed as moving at an instantaneous constant-speed, so that we can treat the entire motion segment by segment, as illustrated in the trajectory of the moving medium. This is the general idea of using a straight line to represent an arc curve in calculus if the segment is rather short. In the moving frame that is affixed to the medium in the $S'$ frame, in which the medium is momentarily at stationary. Therefore, the field observed in a temporarily reference frame $S'$ with those in Lab frame can still be approximately correlated by Minkowski theory, replacing $\boldsymbol{v}_0$ by the total moving speed $\boldsymbol{v}_t$ of the object in Eqs. (13a-f), the momentarily local fields are correlated by:

$$\boldsymbol{E}' \approx \boldsymbol{E} + \boldsymbol{v}_t \times \boldsymbol{B} \tag{16a}$$

$$\boldsymbol{B}' \approx \boldsymbol{B} - \boldsymbol{v}_t \times \boldsymbol{E}/c^2 \tag{16b}$$

$$\boldsymbol{D}' \approx \boldsymbol{D} + \boldsymbol{v}_t \times \boldsymbol{H}/c^2 \tag{16c}$$

$$\boldsymbol{H}' \approx \boldsymbol{H} - \boldsymbol{v}_t \times \boldsymbol{D} \tag{16d}$$

$$\boldsymbol{J}' \approx \boldsymbol{J} - \rho \boldsymbol{v}_t \tag{16e}$$

$$\rho' \approx \rho - \boldsymbol{v}_t \cdot \boldsymbol{J}/c^2 \tag{16f}$$

The theory of Lorentz transformation has been extrapolated to treat the electrodynamics of a moving medium/object by assuming following: 1. The applicability of the Lorentz transformation in medium case, which means the invariance of speed of light; 2. The format-invariance of the Maxwell's equation for medium case. We believe that these are just hypothesis.

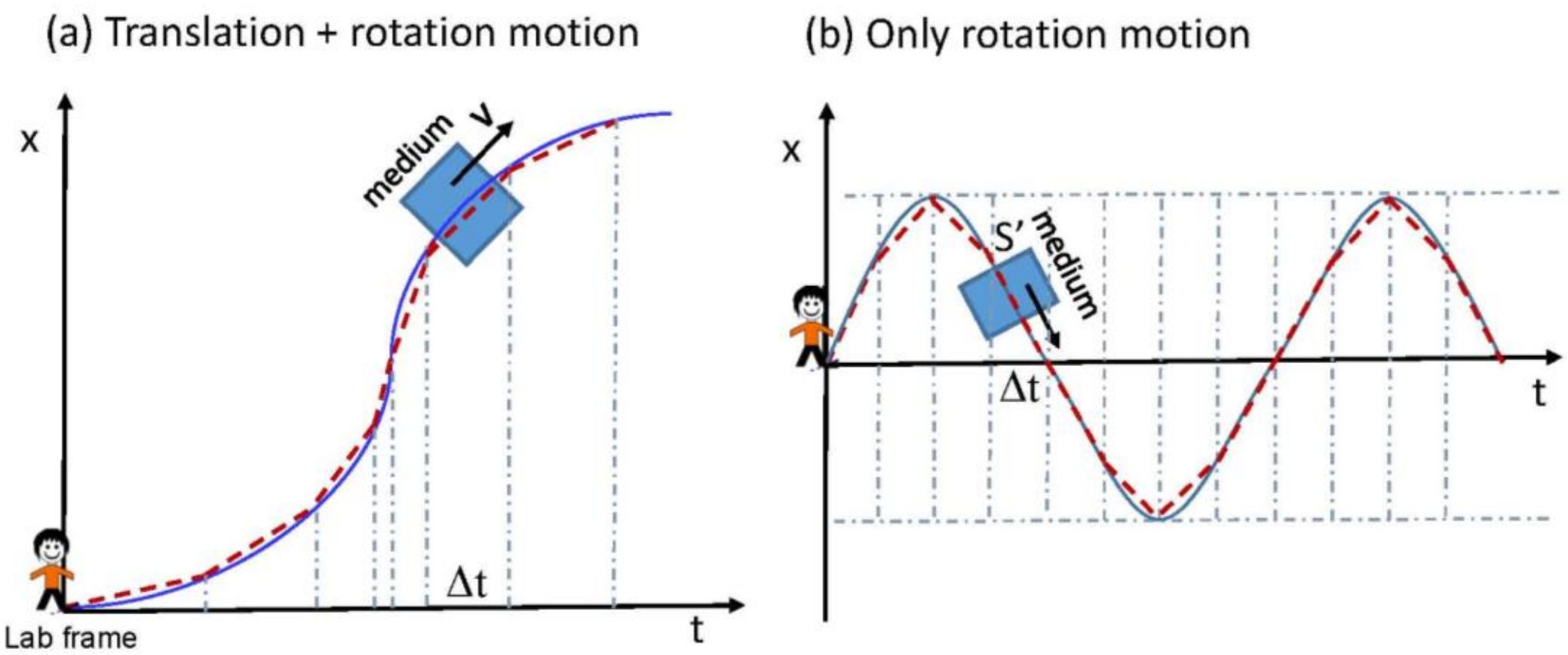

Figure 3. Approach for treating the electrodynamic behavior of an accelerated moving medium by using an integration of segmented path, the movement velocity in each segment can be approximated as a constant so that we can use the constitutive relations derived using the Lorentz transformation for this specific moving velocity. (a) The medium has both translation motion and rotation motion; (b) the medium only has rotation motion.

**4.2 Faraday's law for an accelerated moving medium**

Faraday's law is the fundamental law for electromagnetic generator, which must have external input energy to accelerate the magnets. The derivation of the MEs-f-MDMS was systematically described previously. Now we use the coordination transformation to re-derive the equations. A mathematical identity was used here for the calculations in Eq. (18): for a general functions $\boldsymbol{G}(\boldsymbol{r},t)$, the boundary surface of which moves at a velocity field $\boldsymbol{v}(\boldsymbol{r},t)$ [20], we have

$$\frac{d}{dt}\iint_C \boldsymbol{G}\cdot d\boldsymbol{s} = \iint_C \left[\frac{\partial}{\partial t}\boldsymbol{G} + (\boldsymbol{G}\cdot\nabla)\boldsymbol{v} + (\nabla\cdot\boldsymbol{G})\boldsymbol{v} - (\nabla\cdot\boldsymbol{v})\boldsymbol{G}\right]\cdot d\boldsymbol{s} - \oint_C (\boldsymbol{v}\times\boldsymbol{G})\, d\boldsymbol{L} \tag{17}$$

We first start from the Lenz's law for deriving the Faraday's law, which were derived from experiments. The physics law as shown in Eq. (18) in integral form is usually given in Galilean space-time, so that they can be treated independently. For a closed circuit that contains a large size medium/object in its loop as shown in Fig. 4, the electromotive force around the circuit loop is the reducing rate of the magnetic flux [21]:

$$\xi_{\mathrm{EMF}} = -\frac{d}{dt}\iint_{\boldsymbol{s}(t)} \boldsymbol{B}\cdot d\boldsymbol{s} = -\iint_{\boldsymbol{s}(t)}\left[\frac{\partial}{\partial t}\boldsymbol{B}\cdot d\boldsymbol{s} + \boldsymbol{B}\cdot\frac{\partial}{\partial t}d\boldsymbol{s}\right] = -\iint_{\boldsymbol{s}(t)}\left\{\frac{\partial}{\partial t}\boldsymbol{B} - \nabla\times[\boldsymbol{v}\times\boldsymbol{B}]\right\}\cdot d\boldsymbol{s} \tag{18}$$

where $\boldsymbol{v}$ is the translational moving velocity of the boundary surface $\boldsymbol{s}(t)$ of the medium for calculating the magnetic flux, and it depends only on time, $\boldsymbol{v}(t)$, so that: $(\boldsymbol{G}\cdot\nabla)\boldsymbol{v} = 0$, $(\nabla\cdot\boldsymbol{v})\boldsymbol{G} = 0$, and use $\nabla\cdot\boldsymbol{B} = 0$.

The calculation of the electromotive force can be carried out in the moving $S'$ frame as a circulation flux of the electric field [22],

$$\xi_{\mathrm{EMF}} = \oint_C \boldsymbol{E}'\cdot d\boldsymbol{L} = -\iint_{s(t)}\left\{\frac{\partial}{\partial t}\boldsymbol{B} - \nabla\times[\boldsymbol{v}\times\boldsymbol{B}]\right\}\cdot d\boldsymbol{s} \tag{19}$$

where $\boldsymbol{E}'$ represents the electric field in the rest frame of each segment $d\boldsymbol{L}$ of the path of integration on the circuit that is relative at stationary. The Lorentz force acting on a point charge $q$ in the Lab frame is $\boldsymbol{F} = q\,(\boldsymbol{E}+\boldsymbol{v}_t\times\boldsymbol{B})$, and the force acting on the charge in the moving frame is $\boldsymbol{F}' = q\boldsymbol{E}'$, where $\boldsymbol{v}_t$ is the total moving velocity of the point charge that may be different from the moving velocity $\boldsymbol{v}$ of the circuit loop because of the motion of the point charge inside the large medium depending on its relative rotation/trajectory. If the inertia force is ignored, we may have: $\boldsymbol{F}' \approx \boldsymbol{F}$. Therefore, the local electric field in moving frame is related to that in the Lab frame for the instantaneous inertia frame is [23]

$$\boldsymbol{E}' \approx \boldsymbol{E} + \boldsymbol{v}_t\times\boldsymbol{B}. \tag{20}$$

The total moving velocity $\boldsymbol{v}_t$ of a point charge is a sum of the circuit surface moving velocity $\boldsymbol{v}$ plus the relative motion of the charge inside the large dielectric medium $\boldsymbol{v}_r$. More specifically, $\boldsymbol{v}$ is attributed to be time-dependent so that it can be viewed as a "rigid translation" of the circuit, and $\boldsymbol{v}_r$ is the relative moving velocity of the point charge in the $S'$ frame, which is space- and time-dependent [11,12]

$$\boldsymbol{v}_t = \boldsymbol{v}(t) + \boldsymbol{v}_r(\boldsymbol{r},t) \tag{21}$$

Substituting Eqs. (20) and (21) into Eq. (19), we have

$$\oint_C (\boldsymbol{E} + \boldsymbol{v}_r \times \boldsymbol{B}) \cdot d\boldsymbol{L} = -\iint_s \frac{\partial}{\partial t}\boldsymbol{B} \cdot d\boldsymbol{s} \tag{22}$$

The above integral has two cases to consider along the circuit loop as shown in Fig. 4.

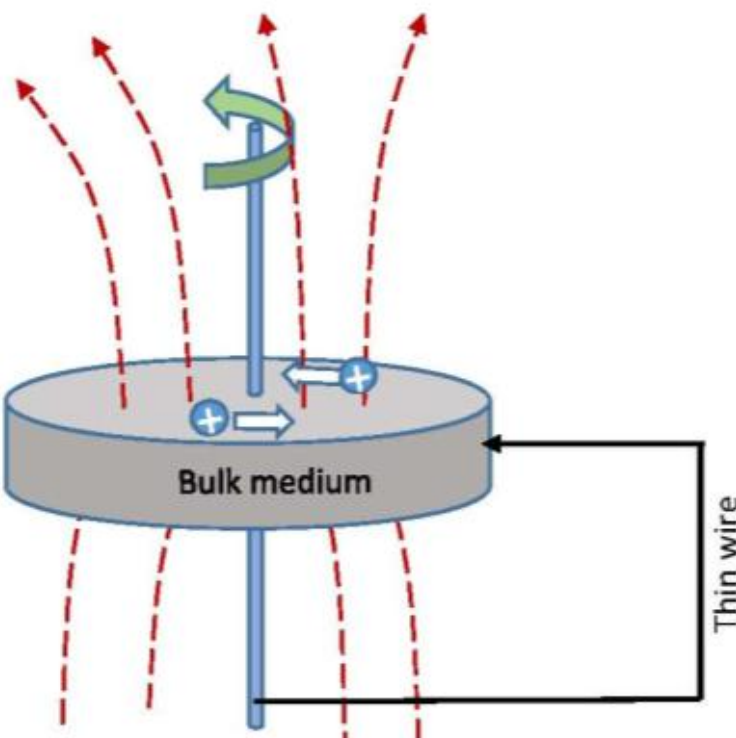

Figure 4. A schematic diagram showing a closed circuit loop that contains a rotating conductive disc with the contacting wire sliding on the surface of the disc. The disc may have surface charges and is rotating in an electromagnetic field. The electromagnetic behavior inside the disc is governed by the MEs-f-MDMS.

The first case is a thin wire case, so that the moving velocity $\boldsymbol{v}_r$ of a point charge inside the wire is parallel to the wire, e.g., $\boldsymbol{v}_r$ is parallel to the integral path $d\boldsymbol{L}$, the term $[\boldsymbol{v}_r \times \boldsymbol{B}\,] \cdot d\boldsymbol{L}$ vanishes automatically. In this case, using the Stokes theorem, the Faraday's law is

$$\nabla \times \boldsymbol{E} = -\frac{\partial}{\partial t}\boldsymbol{B}. \tag{23}$$

This means that the standard form of the Faraday's law in differential form is valid for all thin-wire segments. If the wire is thin enough, it can be considered outside of the medium, in which the classical Maxwell's equations are exact. Theoretically, the thin wire circuit is an imaginary circuit whose shape is arbitrary. If the imaginary circuit can be expanded to any space, the thin circuit assumption holds exactly if the circuit does not intercept with any conductive medium boundary.

The second case is the presence of a large conductive medium, in which the relative velocity of the unit charge is not parallel to the integral path, so that $\oint_C [\boldsymbol{v}_r \times \boldsymbol{B}] \cdot d\boldsymbol{L} \neq 0$. Such cases occur in engineering due to the size and shape of a rotating media. With respect to the distance the electromagnetic fields are considered, the size of the medium may not be negligible. This was first presented by Feynman as the "anti-flux" examples arising from the rotation of a metal disc in a magnetic field. Now we use the Stokes theorem, Eq. (22) becomes [12]

$$\nabla \times (\boldsymbol{E} + \boldsymbol{v}_r \times \boldsymbol{B}) = -\frac{\partial}{\partial t}\boldsymbol{B} \tag{24}$$

The term $\boldsymbol{v}_r \times \boldsymbol{B}$ represents the local Lorentz force of the charge that wonders inside a moving medium. This is a key addition in the equation. One must note that this term was not present in Minkowski's theory even for an accelerated moving object, because the format-invariance was assumed at first place.

### 4.3 The Ampere-Maxwell law

Similarly, the medium motion can be introduced in the Ampere-Maxwell law based on the same logic. In the S' frame, the circulation of the magnetic field $\boldsymbol{H'}$ along a circuit is the total current including conduction current and displacement current, which is given by:

$$\oint_C \boldsymbol{H'} \cdot d\boldsymbol{L} \approx \iint_s \ \boldsymbol{J} \cdot d\boldsymbol{s} + \frac{d}{dt} \iint_s \ \boldsymbol{D} \cdot d\boldsymbol{s} \tag{25}$$

where the $\boldsymbol{H'}$ is the magnetic field on the moving medium in the frame where $d\boldsymbol{L}$ is at rest. The Ampere-Maxwell's law means that the circulation of the magnetic field around its looped edge is the total conductive current through an open surface plus the increasing rate of the electric flux through an open surface (e.g., the displacement current). Using the flux theorem Eq. (17) , assume the boundary movement is a solid translation, $\boldsymbol{v}$ (t), and $\nabla \cdot \boldsymbol{D} \approx \rho$, we have

$$\oint_C \boldsymbol{H'} \cdot d\boldsymbol{L} = \iint_s \ (\boldsymbol{J} + \rho \boldsymbol{v}\,) \cdot d\boldsymbol{s} + \iint_s \ \frac{\partial}{\partial t}\boldsymbol{D} \cdot d\boldsymbol{s} - \oint_C (\,\boldsymbol{v} \times \boldsymbol{D}) \cdot d\boldsymbol{L} \tag{26}$$

Under low-speed approximation, $D \gg |\boldsymbol{v}_t \times \boldsymbol{H}|/c^2$, we have $\boldsymbol{H'} \approx \boldsymbol{H} - \boldsymbol{v}_t \times \boldsymbol{D}$, and using the Stokes theorem, under the first order approximation, we have [11,12]

$$\nabla \times (\boldsymbol{H} - \boldsymbol{v}_r \times \boldsymbol{D}) \approx \boldsymbol{J} + \rho \boldsymbol{v} + \frac{\partial}{\partial t}\boldsymbol{D} \tag{27}$$

Here we only keep the first order terms of velocity in the equation. The terms $\boldsymbol{v}_r \times \boldsymbol{D}$ is the electric field induced magnetization due to media movement. One must note that this term was not present in Minkowski's theory even for an accelerated moving object, because the format-invariance was assumed at first place.

### 4.4 The MEs-f-MDMS

We now put all of the derivations together, the electrodynamic behavior inside an accelerated moving medium can be described by (see Fig. 5) [12,13]

$$\nabla \cdot \boldsymbol{D} \approx \rho \tag{28a}$$

$$\nabla \cdot \boldsymbol{B} \approx 0 \tag{28b}$$

$$\nabla \times (\boldsymbol{E} + \boldsymbol{v}_r \times \boldsymbol{B}) \approx -\frac{\partial}{\partial t}\boldsymbol{B} \tag{28c}$$

$$\nabla\times(\boldsymbol{H}-\boldsymbol{v}_r\times\boldsymbol{D})\approx\boldsymbol{J}+\rho\boldsymbol{v}+\frac{\partial}{\partial t}\boldsymbol{D} \tag{28d}$$

The charge conservation law is:

$$\nabla\cdot(\boldsymbol{J}+\rho\boldsymbol{v})+\frac{\partial}{\partial t}\rho=0 \tag{28e}$$

The Gauss law for electricity and magnetism (Eqs. (28a-b) holds if one ignores the high order terms in velocity. The MEs-f-MDMS describes the force fields coupling among electric-magnetic-mechanical fields. The traditional Maxwell equation considers only the electro-magnetic coupling in inertia reference frame.

The conditions under which Eqs. (28a-d) were derived are: (1) low-speed approximation, and to keep only the first order terms of the moving speed in the equations, and (2) $B \gg v_t E/c^2$ (for Eq. (28b)), and $D \gg v_t H/c^2$ (for Eq. (28d)), so that $\boldsymbol{D}' \approx \boldsymbol{D}$ and $\boldsymbol{B}' \approx \boldsymbol{B}$. Thus, the relativistic corrections made in $\boldsymbol{D}$ and $\boldsymbol{B}$ were ignored. This is the result of Galilean transformation. The two conditions are equivalent to: $\varepsilon\mu/\varepsilon_0\mu_0 >> (v_t/c)^2$, which means that these approximations are excellent in practice.

We now look at two special cases. Outside the medium in free-space, $v_r = 0$, and $v = 0$, the MEs-f-MDMS resumes the standard format of the Maxwell's equations. Alternatively, if an object moves with uniform velocity without acceleration, so that $\boldsymbol{v}_r = 0, \boldsymbol{v} = constant$, the MEs-f-MDMS reduces to the standard format of the Maxwell's equations. This agrees with the Minkowski's approach. Therefore, MEs-f-MDMS is entirely consistent with the Maxwell's equations under considered conditions.

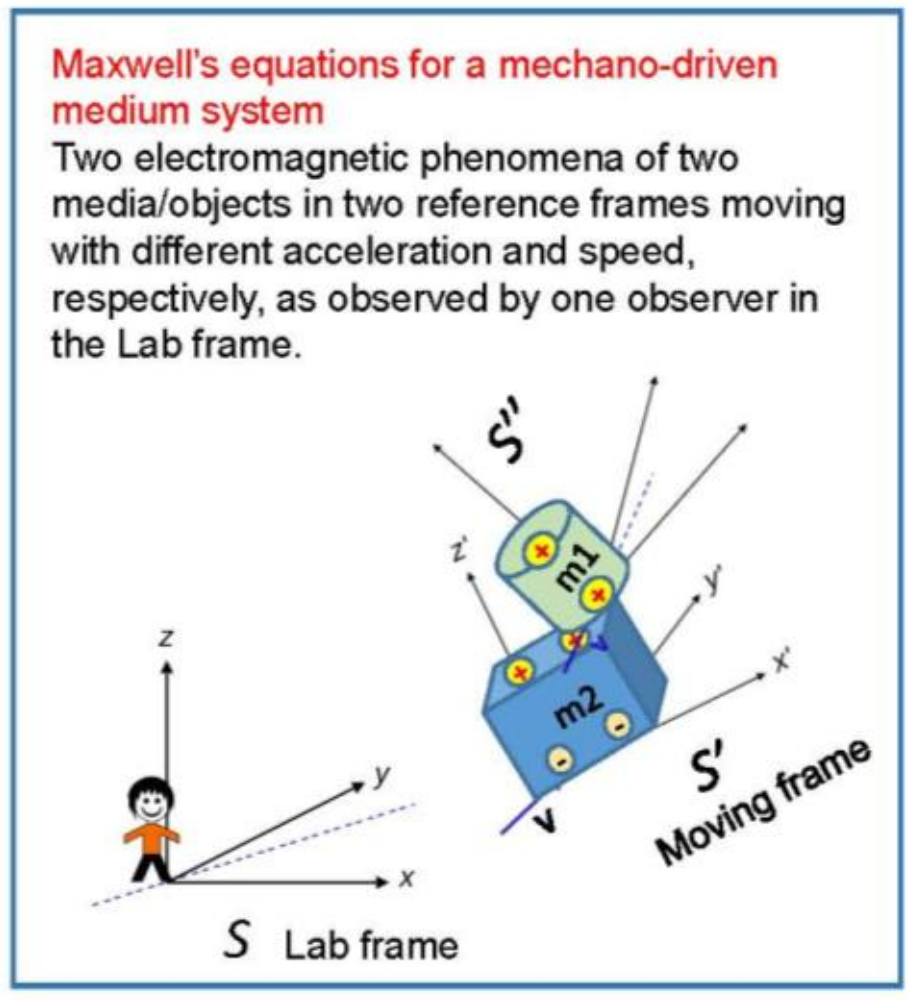

Figure 5. Schematic diagram showing the configuration for deriving the electrodynamics of moving media/objects that may have different movement velocities and accelerations using the MEs-f-MDMS. Media M1 and M2 are affixed to the two moving reference frames $S'$ and $S''$, respectively.

## 5. The constitutive equations for a moving medium that moves with acceleration

We now derive the constitutive equations for an accelerated moving medium. In the moving reference frame that is affixed to the medium, the conventional constitutive relationship in the $S'$ frame are given by Eqs. (14a-b). Replacing $\boldsymbol{v}_0$ by $\boldsymbol{v}_t$, we have the constitutive equations for MEs-f-MDMS:

$$\boldsymbol{D} \approx \varepsilon \boldsymbol{E} + \alpha\varepsilon\mu[\boldsymbol{v}_t \times \boldsymbol{H} - \varepsilon \boldsymbol{v}_t(\boldsymbol{v}_t \cdot \boldsymbol{E})] \approx \varepsilon \boldsymbol{E} + \alpha\varepsilon\mu \boldsymbol{v}_t \times \boldsymbol{H} \tag{29a}$$

$$\boldsymbol{B} \approx \mu \boldsymbol{H} - \alpha\varepsilon\mu[\boldsymbol{v}_t \times \boldsymbol{E} - \mu \boldsymbol{v}_t(\boldsymbol{v}_t \cdot \boldsymbol{H})] \approx \mu \boldsymbol{H} - \alpha\varepsilon\mu\, \boldsymbol{v}_t \times \boldsymbol{E} \tag{29b}$$

Please note that Eqs. (29a-b) are given using the Lorentz transform for free-space, in which the speed of light is assumed to be the speed in free-space rather than that in the medium that is: $c_m = 1/\sqrt{\varepsilon\mu}$ .

## 6. The boundary conditions for MEs-f-MDMS

As elaborated in Section 3, in the moving frame that is affixed to the medium, the corresponding boundary conditions between the media at the two sides of the boundary in the $S'$ frame can be derived by using the integral form of the Maxwell's equations around a small volume or loop [22], and results are given by:

$$\boldsymbol{n}' \cdot (\boldsymbol{D}_2' - \boldsymbol{D}_1') = \sigma' \tag{30a}$$

$$\boldsymbol{n}' \cdot (\boldsymbol{B}_2' - \boldsymbol{B}_1') = 0 \tag{30b}$$

$$\boldsymbol{n}' \times (\boldsymbol{E}_2' - \boldsymbol{E}_1') = 0 \tag{30c}$$

$$\boldsymbol{n}' \times (\boldsymbol{H}_2' - \boldsymbol{H}_1') = \mathbf{K}_s' \tag{30d}$$

where $\boldsymbol{n}'$ is the surface normal direction in the $S'$ frame, $\sigma'$ is the surface free charge density, $\mathbf{K}_s'$ is the surface current density, and the subscripts 1 and 2 represent the two media adjacent to the boundary, as presented in Fig. 4.

However, the boundary conditions for a moving medium may not be derived following the classical approach of applying the integral form of the Maxwell's equations around a small volume of a small loop at the boundary, because the configuration of the boundary varies due to medium motion. We now use the instantaneous coordination transformation approach for deriving the boundary condition using in the differential form of the equations. In the moving reference frame, in which the boundary conditions are given by Eqs. (30a-d), because the medium is instantaneously at stationary. Therefore, we can substitute the field transformation as given in Eq. (16a-d) into (30a-d). To illustrate the configuration clearly, Fig. 5 shows two moving frames $S'$ and $S''$, and there are media m1 and m2 affixed to each, and both media can move with different velocities and may form a boundary. We use the constitutive equations derived under Lorentz transformation to convert the boundary conditions in Lab frame. The boundary conditions between the two media are given by:

$$\boldsymbol{n}\cdot[\boldsymbol{D}_2 - \boldsymbol{D}_1 + (\boldsymbol{v}_{t2}\times\boldsymbol{H}_2 - \boldsymbol{v}_{t1}\times\boldsymbol{H}_1)/c^2] \approx \sigma' \quad (31a)$$

$$\boldsymbol{n}\cdot[\boldsymbol{B}_2 - \boldsymbol{B}_1 - (\boldsymbol{v}_{t2}\times\boldsymbol{E}_2 - \boldsymbol{v}_{t1}\times\boldsymbol{E}_1)/c^2] \approx 0 \quad (31b)$$

$$\boldsymbol{n}\times[\boldsymbol{E}_2 - \boldsymbol{E}_1 + \boldsymbol{v}_{t2}\times\boldsymbol{B}_2 - \boldsymbol{v}_{t1}\times\boldsymbol{B}_1] \approx 0 \quad (31c)$$

$$\boldsymbol{n}\times[\boldsymbol{H}_2 - \boldsymbol{H}_1 - \boldsymbol{v}_{t2}\times\boldsymbol{D}_2 + \boldsymbol{v}_{t1}\times\boldsymbol{D}_1] \approx \mathbf{K}_s' \quad (31d)$$

and

$$\mathbf{K}_s' \approx \mathbf{K}_{s2} + \mathbf{K}_{s1} - (\sigma_2\boldsymbol{v}_{t2} + \sigma_1\boldsymbol{v}_{t1}) \quad (31e)$$

$$\sigma' \approx \sigma_2 + \sigma_1 - (\boldsymbol{v}_{t2}\cdot\boldsymbol{J}_2 + \boldsymbol{v}_{t1}\cdot\boldsymbol{J}_1)/c^2 \quad (31f)$$

where $\boldsymbol{n}$ is the surface normal direction converted from $\boldsymbol{n}'$ using Lorentz transformation; $\mathbf{K}_s$ is the surface current density by considering the contribution from the media adjacent to the boundary; and the subscripts 1 and 2 represent the corresponding quantities inside the dielectric media at the two sides of the boundary. The compete derivation of the constitutive relations, the basic equations and the boundary conditions make the numerical calculation possible.

## 7. **The first order approximated result of MEs-f-MDMS**

The following discussions are given by assuming that we only keep the first order terms of $\boldsymbol{v}_t$ in the constitutive relations, the MEs-f-MDMS and boundary conditions. From Eq. (29a-b), we have the first order approximated result:

$$\boldsymbol{D} \approx \varepsilon\boldsymbol{E} + \alpha\varepsilon\mu\, \boldsymbol{v}_t\times\boldsymbol{H} \quad (32b)$$

$$\boldsymbol{B} \approx \mu\boldsymbol{H} - \alpha\varepsilon\mu\, \boldsymbol{v}_t\times\boldsymbol{E} \quad (32a)$$

Both speeds of light in free-space and in medium appear in the equation. This is a consequence of using the Lorentz transformation for free-space to treat the electromagnetic behavior of a moving dielectric medium case. The question is how to expand Lorentz transformation in the space there are finite size media with the presence of boundaries.

Substituting Eqs. (32a-b) into (28a-d), the MEs-f-MDMS inside the dielectric medium is

$$\varepsilon\,\nabla\cdot(\boldsymbol{E} + \alpha\mu\, \boldsymbol{v}_t\times\boldsymbol{H}) \approx \rho \quad (33a)$$

$$\mu\,\nabla\cdot(\boldsymbol{H} - \alpha\varepsilon\, \boldsymbol{v}_t\times\boldsymbol{E}) \approx 0 \quad (33b)$$

$$\nabla\times(\boldsymbol{E} + \mu\, \boldsymbol{v}_r\times\boldsymbol{H}) \approx -\mu\frac{\partial}{\partial t}(\boldsymbol{H} - \alpha\varepsilon\, \boldsymbol{v}_t\times\boldsymbol{E}) \quad (33c)$$

$$\nabla\times(\boldsymbol{H}-\varepsilon\,\boldsymbol{v}_r\times\boldsymbol{E})\approx\boldsymbol{J}+\rho\boldsymbol{v}+\varepsilon\frac{\partial}{\partial t}(\boldsymbol{E}+\alpha\mu\,\boldsymbol{v}_t\times\boldsymbol{H}) \tag{33d}$$

The solutions of Eq. (33a-d) can be received using the perturbation theory as presented in [13]. The corresponding boundary conditions under first order approximation are given by Eq. (26a-d):

$$\boldsymbol{n}\cdot[\varepsilon_2\boldsymbol{E}_2-\varepsilon_1\boldsymbol{E}_1+\varepsilon_2\mu_2\,\boldsymbol{v}_{t2}\times\boldsymbol{H}_2-\varepsilon_1\mu_1\boldsymbol{v}_{t1}\times\boldsymbol{H}_1]\approx\sigma' \tag{34a}$$

$$\boldsymbol{n}\cdot[\mu_2\boldsymbol{H}_2-\mu_1\boldsymbol{H}_1-\varepsilon_2\mu_2\,\boldsymbol{v}_{t2}\times\boldsymbol{E}_2+\varepsilon_1\mu_1\boldsymbol{v}_{t1}\times\boldsymbol{E}_1]\approx 0 \tag{34b}$$

$$\boldsymbol{n}\times[\boldsymbol{E}_2-\boldsymbol{E}_1+\mu_2\,\boldsymbol{v}_{t2}\times\boldsymbol{H}_2-\mu_1\boldsymbol{v}_{t1}\times\boldsymbol{H}_1]\approx 0 \tag{34c}$$

$$\boldsymbol{n}\times[\boldsymbol{H}_2-\boldsymbol{H}_1-\varepsilon_2\,\boldsymbol{v}_{t2}\times\boldsymbol{E}_2+\varepsilon_1\boldsymbol{v}_{t1}\times\boldsymbol{E}_1]\approx\mathbf{K}_s' \tag{34d}$$

## 8. The MEs-f-MDMS for a rotating medium in free-space

Now we consider a case that a medium is rotating in free-space, such as a disc-like rotating object. The electromagnetic field inside the medium are governed by the MEs-f-MDMS as given by Eqs. (33a-d) by taking $\boldsymbol{v}=0$, and $\boldsymbol{v}_t=\boldsymbol{v}_r$:

$$\varepsilon\,\nabla\cdot(\boldsymbol{E}+\alpha\mu\,\boldsymbol{v}_r\times\boldsymbol{H})\approx\rho \tag{35a}$$

$$\mu\,\nabla\cdot(\boldsymbol{H}-\alpha\varepsilon\,\boldsymbol{v}_r\times\boldsymbol{E})\approx 0 \tag{35b}$$

$$\nabla\times(\boldsymbol{E}+\mu\,\boldsymbol{v}_r\times\boldsymbol{H})\approx-\mu\frac{\partial}{\partial t}(\boldsymbol{H}-\alpha\varepsilon\,\boldsymbol{v}_r\times\boldsymbol{E}) \tag{35c}$$

$$\nabla\times(\boldsymbol{H}-\varepsilon\,\boldsymbol{v}_r\times\boldsymbol{E})\approx\boldsymbol{J}+\varepsilon\frac{\partial}{\partial t}(\boldsymbol{E}+\alpha\mu\,\boldsymbol{v}_r\times\boldsymbol{H}) \tag{35d}$$

The solution of Eqs. (35a-d) will be presented in Section 10. Outside the dielectric medium, the electromagnetic fields are governed by:

$$\varepsilon_0\,\nabla\cdot\boldsymbol{E}=\rho \tag{36a}$$

$$\mu_0\,\nabla\cdot\boldsymbol{H}=0 \tag{36b}$$

$$\nabla\times\boldsymbol{E}=-\mu_0\frac{\partial}{\partial t}\boldsymbol{H} \tag{36c}$$

$$\nabla\times\boldsymbol{H}=\boldsymbol{J}+\varepsilon_0\frac{\partial}{\partial t}\boldsymbol{E} \tag{36d}$$

The corresponding boundary conditions Eq. (34a-d) become:

$$\boldsymbol{n}\cdot[\varepsilon_2\boldsymbol{E}_2-\varepsilon_0\boldsymbol{E}_1+\varepsilon_2\mu_2\,\boldsymbol{v}_{r2}\times\boldsymbol{H}_2]\approx\sigma' \tag{37a}$$

$$\boldsymbol{n}\cdot[\mu_2\boldsymbol{H}_2-\mu_0\boldsymbol{H}_1-\varepsilon_2\mu_2\,\boldsymbol{v}_{r2}\times\boldsymbol{E}_2]\approx 0 \tag{37b}$$

$$\boldsymbol{n}\times[\boldsymbol{E}_2-\boldsymbol{E}_1+\mu_2\,\boldsymbol{v}_{r2}\times\boldsymbol{H}_2]\approx 0 \tag{37c}$$

$$\boldsymbol{n}\times[\boldsymbol{H}_2-\boldsymbol{H}_1-\varepsilon_2\,\boldsymbol{v}_{r2}\times\boldsymbol{E}_2]\approx\mathbf{K}_s' \tag{37d}$$

## 9. The MEs-f-MDMS for high refraction index medium with rotation motion

Now we consider a case that the dielectric permittivity of the medium is rather large, so that $\varepsilon\mu \gg \varepsilon_0\mu_0$, such as the case for high dielectric materials and permanent magnets, $\alpha \approx 1$, and $\boldsymbol{v} = 0$, the MEs-f-MDMS as shown in Eqs. (35a-d) are simplified as [23]:

$$\varepsilon \nabla \cdot \boldsymbol{E}_{eff} \approx \rho \tag{38a}$$

$$\mu \nabla \cdot \boldsymbol{H}_{eff} \approx 0 \tag{38b}$$

$$\nabla \times \boldsymbol{E}_{eff} \approx -\mu \frac{\partial}{\partial t} \boldsymbol{H}_{eff} \tag{38c}$$

$$\nabla \times \boldsymbol{H}_{eff} \approx \boldsymbol{J} + \varepsilon \frac{\partial}{\partial t} \boldsymbol{E}_{eff} \tag{38d}$$

where the effective fields are defined by:

$$\boldsymbol{E}_{eff} = \boldsymbol{E} + \mu \boldsymbol{v}_r \times \boldsymbol{H} \tag{39a}$$

$$\boldsymbol{H}_{eff} = \boldsymbol{H} - \varepsilon \boldsymbol{v}_r \times \boldsymbol{E} \tag{39b}$$

Equations (38a-d) are identical to the classical Maxwell equations except that the electric field $\boldsymbol{E}$ and magnetic field $\boldsymbol{H}$ are replaced by the effective fields $\boldsymbol{E}_{eff}$ and $\boldsymbol{H}_{eff}$, respectively. The electrodynamic behavior as described by the effective fields have been described in details previously. The condition for the effective medium to be valid is $\varepsilon\mu \gg \varepsilon_0\mu_0$, such as permanent magnetic materials and ferroelectric or piezoelectric materials.

The corresponding boundary conditions under first order approximation are given by:

$$\boldsymbol{n} \cdot [\varepsilon_2 \boldsymbol{E}_{\text{eff},2} - \varepsilon_1 \boldsymbol{E}_{\text{eff},1}] \approx \sigma' \tag{40a}$$

$$\boldsymbol{n} \cdot [\mu_2 \boldsymbol{H}_{\text{eff},2} - \mu_1 \boldsymbol{H}_{\text{eff},1}] \approx 0 \tag{40b}$$

$$\boldsymbol{n} \times [\boldsymbol{E}_{\text{eff},2} - \boldsymbol{E}_{\text{eff},1}] \approx 0 \tag{40c}$$

$$\boldsymbol{n} \times [\boldsymbol{H}_{\text{eff},2} - \boldsymbol{H}_{\text{eff},1}] \approx \mathbf{K}_s' \tag{40d}$$

The condition of $\varepsilon\mu \gg \varepsilon_0\mu_0$ applies to high refraction or permanent magnetic materials. For general materials, such condition may not be strongly satisfied. The key test is to compare with experimental observations. One may include high orders terms in Eqs. (40a-d) if needed.

The condition of $\varepsilon\mu \gg \varepsilon_0\mu_0$ is equivalent to the total reflection of light inside a high refraction medium, and it most likely to be trapped in the medium. This means that the electromagnetic emission generated by the medium motion inside the medium may not "escape the medium" owing to total reflection, so that one may hardly detect the signals outside of the medium. Therefore, we may built an electromagnetic fiber that made of high dielectric or magnetic materials, so that the low-frequency electromagnetic wave can be transported to a long distance, in analogy to the optical fiber. This can be an approach for transmitting electromagnetic waves for special applications such as under water.

## 10. Solution of the MEs-f-MDMS for a medium without translation motion

We now lay out the analytical solution for Equations (35a-d). We first define two quantities:

$$\boldsymbol{\mathfrak{E}} = \boldsymbol{E} + \alpha\mu\, \boldsymbol{v}_r \times \boldsymbol{H} \tag{41a}$$

$$\boldsymbol{\mathcal{H}} = \boldsymbol{H} - \alpha\varepsilon\, \boldsymbol{v}_r \times \boldsymbol{E} \tag{41b}$$

Using which, Eqs. (35a-d) are written as

$$\varepsilon\, \nabla \cdot \boldsymbol{\mathfrak{E}} \approx \rho \tag{42a}$$

$$\mu\, \nabla \cdot \boldsymbol{\mathcal{H}} \approx 0 \tag{42b}$$

$$\nabla\times \boldsymbol{\mathfrak{E}} \approx \boldsymbol{P}_M - \mu\frac{\partial}{\partial t}\boldsymbol{\mathcal{H}} \tag{42c}$$

$$\nabla\times \boldsymbol{\mathcal{H}} \approx \boldsymbol{P}_{\mathrm{E}} + \varepsilon\frac{\partial}{\partial t}\boldsymbol{\mathfrak{E}} \tag{42d}$$

Where we have defined two quantities, each of which is a function of the fields $\boldsymbol{E}$ and $\boldsymbol{H}$ and is related to the medium moving velocity:

$$\boldsymbol{P}_{\mathrm{M}} = \boldsymbol{P}_{\mathrm{M}}(\boldsymbol{E}, \boldsymbol{H}) = (\alpha - 1)\, \nabla\times (\boldsymbol{v}_r \times\boldsymbol{H}) \tag{43a}$$

$$\boldsymbol{P}_{\mathrm{E}} = \boldsymbol{P}_{\mathrm{E}}(\boldsymbol{E}, \boldsymbol{H}) = \boldsymbol{J} + \rho\boldsymbol{v} - (\alpha - 1)\, \nabla\times (\boldsymbol{v}_r \times\boldsymbol{E}) \tag{43b}$$

Such notation is taken in order to solve the equations using perturbation theory and interaction method. By applying $\nabla\times$ on both Eq. (42c-d), we have:

$$-\nabla^2\boldsymbol{\mathfrak{E}} + \mu\varepsilon\frac{\partial^2}{\partial t^2}\boldsymbol{\mathfrak{E}} \approx \nabla\times\boldsymbol{P}_{\mathrm{M}} - \nabla\rho/\varepsilon - \mu\frac{\partial}{\partial t}\boldsymbol{P}_E \tag{44a}$$

$$-\nabla^2\boldsymbol{\mathcal{H}} + \mu\varepsilon\frac{\partial^2}{\partial t^2}\boldsymbol{\mathcal{H}} \approx \nabla\times\boldsymbol{P}_{\mathrm{E}} + \varepsilon\frac{\partial}{\partial t}\boldsymbol{P}_M \tag{44b}$$

These Holmhertz equations can be solved using the iteration method as reported previously [13].

By applying $\boldsymbol{v}_r\cdot$ and $\boldsymbol{v}_r\times$, respectively, on Eqs. (41a-b), we can derive the electric and magnetic fields as:

$$\boldsymbol{H} = \frac{1}{1-\alpha^2\varepsilon\mu {v_r}^2}\left[\boldsymbol{\mathcal{H}} + \alpha\varepsilon\, \boldsymbol{v}_r \times\boldsymbol{\mathfrak{E}} - \alpha^2\varepsilon\mu\, \boldsymbol{v}_r(\boldsymbol{v}_r \cdot \boldsymbol{\mathcal{H}})\right] \approx \boldsymbol{\mathcal{H}} + \alpha\varepsilon\, \boldsymbol{v}_r \times\boldsymbol{\mathfrak{E}} \tag{45b}$$

$$\boldsymbol{E} = \frac{1}{1-\alpha^2\varepsilon\mu {v_r}^2}\left[\boldsymbol{\mathfrak{E}} - \alpha\varepsilon\, \boldsymbol{v}_r \times\boldsymbol{\mathcal{H}} - \alpha^2\varepsilon\mu\, \boldsymbol{v}_r(\boldsymbol{v}_r \cdot \boldsymbol{\mathfrak{E}})\right] \approx \boldsymbol{\mathfrak{E}} - \mu\varepsilon\, \boldsymbol{v}_r \times\boldsymbol{\mathcal{H}} \tag{45b}$$

The corresponding boundary conditions under the first order approximation are:

$$\boldsymbol{n} \cdot [\varepsilon_2\boldsymbol{E}_2 - \varepsilon_0\boldsymbol{E}_1 + \varepsilon_2\mu_2\, \boldsymbol{v}_{r2} \times\boldsymbol{H}_2] \approx \sigma' \tag{46a}$$

$$\boldsymbol{n}\cdot[\mu_2\boldsymbol{H}_2\ -\mu_0\boldsymbol{H}_1-\ \varepsilon_2\mu_2\ \boldsymbol{v}_{r2}\times\boldsymbol{E}_2]\approx 0 \tag{46b}$$

$$\boldsymbol{n}\times[\boldsymbol{E}_2\ -\boldsymbol{E}_1+\mu_2\ \boldsymbol{v}_{r2}\times\boldsymbol{H}_2]\ \approx 0 \tag{46c}$$

$$\boldsymbol{n}\times[\boldsymbol{H}_2\ -\boldsymbol{H}_1-\varepsilon_2\ \boldsymbol{v}_{r2}\times\boldsymbol{E}_2]\approx \mathbf{K}_s{}' \tag{46d}$$

## 11. **Conclusions**

The Minkowski theory was developed using the Lorentz transformation for a uniformly moving object by constructing the Maxwell's equations to be format-invariant in two inertia reference frames. In this paper, under the instantaneous constant-speed approximation, we have expanded the Minkowski's theory to derived the constitutive equations and boundary conditions for dielectric medium that has accelerated translation and rotation motion and even shape deformation, e.g., $\boldsymbol{v}_t(\boldsymbol{r},t)$. Using the Lorentz transformation under low-speed approximation, we have developed the constitutive equations and boundary conditions for the Maxwell's equations for a mechano-driven medium system (MEs-f-MDMS), so that the full theory for numerical calculations are thus completed.

The Minkowski theory was derived using Lorentz transformation, while the MEs-f-MDMS were derived using Galilean transformation. However, under low-speed and in the experimental environment one is interested in, the two transformations are equivalent. Therefore, a nice coupling of the Minkowski theory with the MEs-f-MDMS provides a systematic theory for treating the electrodynamics of accelerated moving objects.

Figure 6 shows a summary on the main theoretical results we have achieved. For the cases of (a) stationary medium or in free-space, the standard Maxwell's equations should be used. In case (b), the medium is a solid object and its motion is a solid translation without rotation, e.g., $\boldsymbol{v}_t=\ \boldsymbol{v}(t)$, the standard Maxwell's equations should be used except an additional current term $\rho\boldsymbol{v}$ is added. While for the case (c) that the object has both translation and rotation motion, $\boldsymbol{v}_t(\boldsymbol{r},t)$, which depends on both space and time, the ME-f-MDMS should be used. The constitutive equations and the boundary conditions remain the same for cases (b) and (c). In a mechano-driven media system, mechanical action is not merely passive motion through space but an active source of electromagnetic generation and modulation. Our theory makes it possible to treat the electrodynamics by including medium deformation, strain gradients, rotational dynamics, and interfacial contact processes in the full calculation.

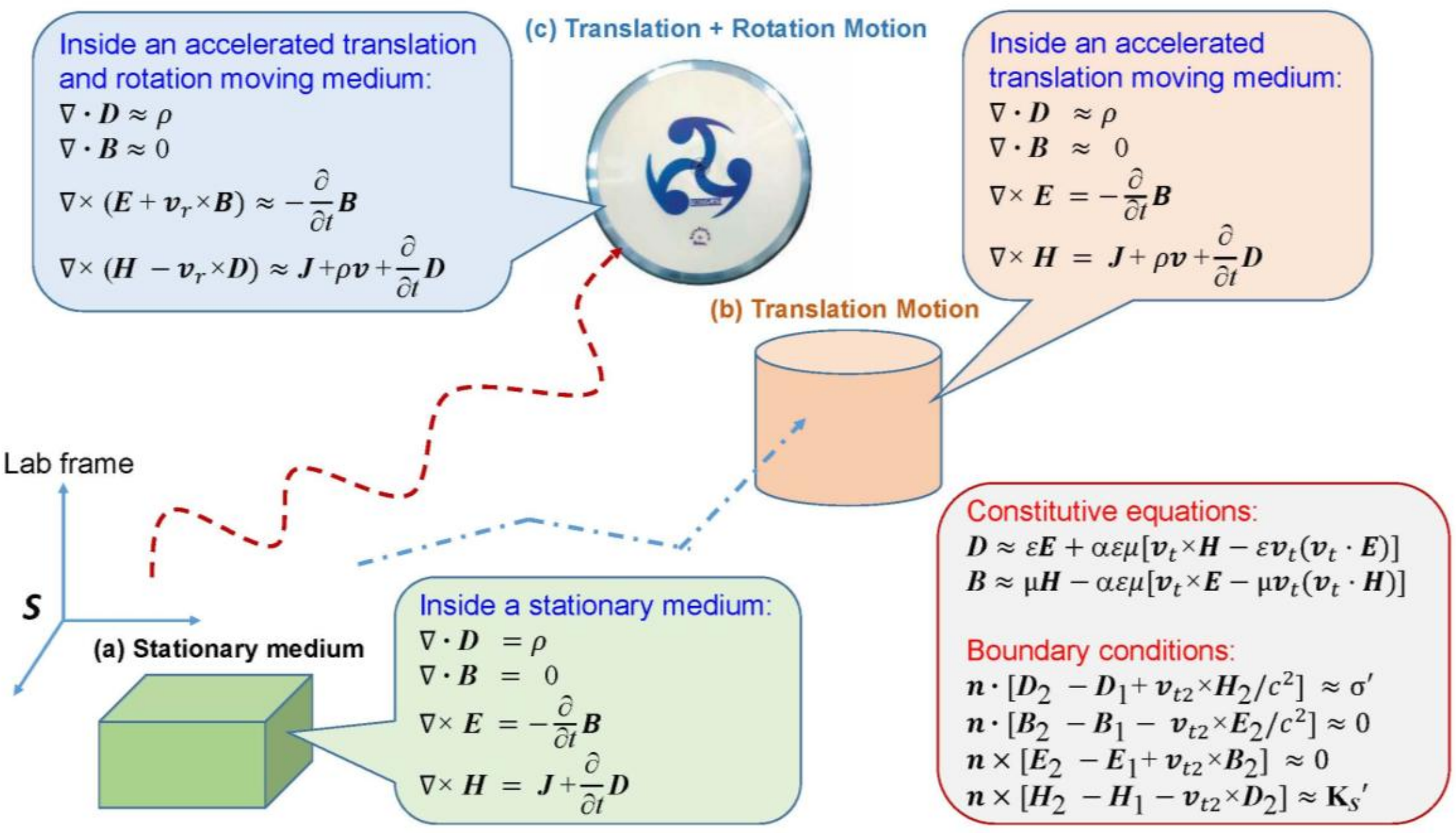

Figure 6. A summary on the theoretical schemes developed for quantitative description of the electrodynamics of various moving dielectric media in Lab reference frame.

**Disclosure statement:** No potential conflict of interest was reported by the author(s).